\documentclass[11pt,a4paper]{article}
\newif\ifphonelayout
\phonelayoutfalse
\usepackage[utf8]{inputenc}
\usepackage[T1]{fontenc}
\usepackage{lmodern}
\ifphonelayout
  \usepackage[a4paper,textwidth=90.3mm,top=22mm,bottom=23mm,centering]{geometry}
\else
  \usepackage[a4paper,margin=25mm]{geometry}
\fi
\usepackage{amsmath,amssymb,graphicx,xcolor,setspace}
\usepackage[super,numbers,sort&compress]{natbib}
\usepackage{booktabs,tabularx,array}
\usepackage[font=small,labelfont=bf]{caption}
\usepackage{subcaption}
\usepackage{tikz}
\usetikzlibrary{arrows.meta,positioning,calc}
\usepackage{microtype,xurl,changepage,calc,needspace,titlesec}
\usepackage[section]{placeins}
\usepackage[hidelinks]{hyperref}
\hypersetup{pdftitle={Causal discernment and the
direction of clinical research},pdfauthor={Carlo Berzuini and Luisa Bernardinelli}}
\newcommand{\indep}{\mathrel{\perp\!\!\!\perp}}

\newcolumntype{L}[1]{>{\raggedright\arraybackslash}p{#1}}
\newcolumntype{Y}{>{\raggedright\arraybackslash}X}
\newlength{\widewidth}
\newlength{\wideoffset}
\newenvironment{widefigure}[1][tbp]{\begin{figure}[#1]\begin{adjustwidth}{-\wideoffset}{-\wideoffset}\centering}{\end{adjustwidth}\end{figure}}
\newenvironment{widetable}[1][tbp]{\begin{table}[#1]\begin{adjustwidth}{-\wideoffset}{-\wideoffset}\centering}{\end{adjustwidth}\end{table}}
\renewcommand{\arraystretch}{1.17}
\titleformat{\section}{\normalfont\Large\bfseries\raggedright\hyphenpenalty=10000}{\thesection}{0.7em}{}
\titleformat{\subsection}{\normalfont\large\bfseries\raggedright\hyphenpenalty=10000}{\thesubsection}{0.7em}{}
\titleformat{\subsubsection}{\normalfont\normalsize\bfseries\raggedright\hyphenpenalty=10000}{\thesubsubsection}{0.7em}{}
\renewenvironment{abstract}{\section*{Abstract}}{}
\definecolor{nodefill}{RGB}{219,234,247}
\definecolor{noderule}{RGB}{60,96,130}
\newcommand{\gfont}{\sffamily\fontsize{11.5}{13.5}\selectfont}
\tikzset{
 cnode/.style={draw=noderule,line width=.65pt,rounded corners=3pt,
  fill=nodefill,inner sep=4pt,minimum width=8mm,minimum height=9mm,
  align=center,font=\gfont},
 ar/.style={-{Latex[length=2.5mm]},line width=1pt},
 bd/.style={ar,densely dashed},
 lnode/.style={cnode,densely dashed,fill=white},
 unode/.style={cnode,densely dotted,fill=white}
}
\definecolor{route}{RGB}{40,85,124}
\definecolor{muted}{RGB}{150,157,163}

\tikzset{
  exnode/.style={
    draw=route, fill=blue!5, rounded corners=2pt,
    minimum width=7mm, minimum height=7mm,
    inner sep=2pt, font=\normalsize
  },
  exmuted/.style={
    exnode, draw=muted, text=muted, fill=white
  },
  exar/.style={
    -{Latex[length=2mm]}, draw=route, line width=1.1pt
  },
  exback/.style={
    -{Latex[length=1.7mm]}, draw=muted, line width=.55pt
  }
}

\newcommand{\exnodes}[1]{%
  \node[exnode] (h) at (0,1.2) {$H$};
  \node[exnode] (p) at (0,-1.2) {$P$};
  \node[exnode] (r) at (2,-.6) {$R$};
  \node[exnode] (l) at (3.9,-.6) {$L$};
  \node[#1]     (w) at (2,1.5) {$W$};
  \node[exnode] (y) at (5.8,.6) {$Y_D$};
}
\title{\textbf{Causal discernment and the direction of clinical research}}
\author{Carlo Berzuini$^1$ and Luisa Bernardinelli$^2$}
\date{}
\makeatletter
\renewcommand{\@maketitle}{%
  \begin{center}
  {\LARGE\@title\par}\vspace{5mm}
  {\large\@author\par}
  \end{center}\vspace{1mm}}
\makeatother
\begin{document}
\maketitle
\begin{center}\small
$^1$Centre for Biostatistics, The University of Manchester, United Kingdom\\[3pt]
$^2$Department of Brain and Behavioral Sciences, The University of Pavia, Italy\\[6pt]
\textbf{Correspondence:} Carlo Berzuini. Email: carlo.berzuini@manchester.ac.uk\\[3pt]
\textbf{ORCID:} Carlo Berzuini 0000-0001-6056-0489; Luisa Bernardinelli 0000-0002-7551-6664
\end{center}

\textbf{Running title:} Causal discernment in clinical research

\begin{abstract}
\textbf{Rationale, aims and objectives.}
A clinical trial can establish an intervention's effects while leaving
unresolved which causal pathways produce them, which matters to
subsequent research and treatment. We call \emph{causal
discernment} the practice of identifying competing causal
explanations that fit the evidence but would lead to different
decisions, and seeking evidence that distinguishes them across
a programme of studies.

\textbf{Method.}
We combine causal graphs with accompanying prose, organised around
four questions: whether the intervention delivers the intended
target change, through which routes it produces benefit and harm,
which causal relations remain stable across regimes, and whether
unresolved pathways could change the decision.
Beyond their established use in identification, causal graphs support
plausible reasoning: with quantitative assumptions, they show which
findings and further comparisons distinguish competing explanations,
even where the effects are not identified. The prose states what the
graphs leave out: the clinical meaning of the variables, the size and
timing of effects and the supporting evidence. We develop the
approach through conceptual analysis, hypothetical calculations
and two examples: CAST, which illustrates the distinction between
an agent's total effect and its target's contribution, and lipid-lowering therapy.

\textbf{Substantive results.}
In the lipid example, concordant coronary associations
for HMGCR and PCSK9 perturbations support LDL mediation
over target-specific explanations that predict unequal
effects per unit of LDL reduction. They do not distinguish
LDL mediation from a shared receptor pathway.
The diabetes associations likewise leave LDL lowering,
receptor-mediated uptake and weight gain as possible
explanations. An illustrative causal record specifies
a genetic comparison and the assumptions under which
it could help distinguish these explanations.

\textbf{Conclusion.}
The causal record makes the competing explanations, their
supporting evidence and assumptions explicit, and identifies
what each proposed design could resolve. Prepared before a
design or recommendation is adopted, it opens the reasoning
to scrutiny, and the results are interpreted as evidence among
the explanations it states. Its added value over ordinary
rationale review remains to be assessed.
\end{abstract}


\bigskip

\section{Introduction}
\label{sec:intro}

A randomised trial can establish an intervention's overall
effects on clinical outcomes while leaving unresolved causal
questions that matter to subsequent research or treatment
decisions. A favourable balance
of benefits and harms
may justify treatment in the population and conditions studied
without identifying the causal pathways responsible. Yet different causal explanations of the same result may suggest different choices about which agent to test next, how to avoid an adverse effect or whether to recommend treatment in another care setting.
An unfavourable result can likewise leave open whether the intervention failed to change its target, whether changing the target brought no clinical benefit, or whether a benefit was outweighed by other effects.

We call \emph{causal discernment} the practice of identifying competing causal structures that fit the available evidence but suggest different next studies or treatment choices, and seeking evidence that could distinguish them. In designing a study, investigators are asked to formalise the question it is meant to answer. We propose to formalise the {\em reasoning} by which that question is chosen, and with it the direction of a research programme. Which causal structures are compatible with the evidence so far? Under what conditions would they predict different observable outcomes, and what study could distinguish them? Answering these questions may even change the question the programme pursues. Making this reasoning explicit allows others to examine how the proposed design follows from the evidence and the causal assumptions.

This aim we pursue by combining causal graphs with 
prose to formulate competing causal explanations around
four questions relevant to research and treatment decisions:
whether the intervention produces the intended target change; through which pathways it produces clinical benefits and harms; which causal relations can be assumed to hold across the settings or modes of intervention being compared; and whether unresolved or omitted pathways could change the research or treatment choice.

We use graphs both to ask whether a causal effect can be identified under stated assumptions and to work out what evidence could distinguish competing explanations. Identification asks whether the available data distributions, together with the assumptions represented by a graph, determine a unique value for a specified causal effect.
Pearl's do-calculus provides rules for deriving such results
without specifying parametric forms for the causal relations.\cite{pearl1995}
Discrimination works across competing graphs and asks which observations, or which comparisons between regimes, would distinguish one structure from the others. Discrimination we approach
via plausible reasoning in P\'olya's sense: assessing how observations would alter the credibility of a structure relative to its rivals.\cite{polya1954}
Graphs, together with assumptions about the size and timing of effects, help specify what we would expect to observe under each explanation. The accompanying prose explains why these assumptions are considered reasonable.
Graphs and prose together should also show whether a feasible study could detect the differences that matter to a decision. Experimental and observational evidence can be combined to this end if the assumptions connecting their regimes are justified.

We propose a \emph{causal record}, dated and with named authors, that states the decision to be made, the competing explanations, the evidence supporting them and the findings that would prompt their revision. Prepared before a design or recommendation is adopted, the record makes the reasoning available for scrutiny and preserves the grounds for subsequent changes. Recording the explanations before the study results are known limits the scope for choosing an explanation simply because it fits those results.

Section~\ref{sec:complementary} develops the approach through
lipid-lowering interventions. We use graphs to examine what trial and genetic evidence suggest about the pathways responsible for cardiovascular benefit and diabetes risk, and what remains uncertain.
An illustrative causal record connects the evidence and
remaining uncertainties to a genetic comparison that
could separate two explanations of the diabetes risk.
Other examples address delivery, comparisons between therapeutic agents
and transfer across care settings. Hypothetical calculations
show how assumptions about stability and additional agent
effects affect inference and study choice.

\bigskip

\section{Four questions for causal interpretation}
\label{sec:four-questions}

\medskip


\subsection{Agents, targets and the comparison of interest}

We use \emph{agent} broadly to mean a specified intervention:
a drug regimen, a procedure or an arrangement of care,
including its rules of administration.
The \emph{target} is the biological variable or clinical process
that the intervention is intended to change. Its profile includes
the amount, timing and duration of that change.
An agent-level question concerns the consequences of administering
a particular intervention; a target-level question concerns the
consequences of changing the target profile.

An agent-level question may concern
the outcome effect of assignment to a treatment
strategy or the effect of 
enforcing full adherence to it.\cite{hernan2026estimands}
Neither effect, by itself, identifies how much of the
intervention's effect is mediated by the target.
Nor does it establish the effect of changing the target
profile through another intervention.

The Cardiac Arrhythmia Suppression Trial (CAST) illustrates
this. Frequent ventricular premature
beats after myocardial infarction 
had been found to predict sudden death.
Their suppression appeared
to be a plausible therapeutic aim.
Yet encainide and flecainide suppressed ectopy while increasing
mortality relative to placebo.\cite{cast1989}
Only patients whose ectopy had been suppressed during an initial treatment phase were randomised to compare
continued drug use with withdrawal
to placebo. Because any difference in mortality combined the effect of suppression with every other effect of the drug, the comparison could not tell the two effects apart.

How much of the effect of these drugs passed through suppression
is a mediation question, requiring assumptions that randomisation
does not provide.\cite{vanderweele2016}
Even if achievable, the answer would be restricted to the particular drugs on test, and would leave open the target-level question whether a comparable suppression profile would improve survival if achieved through another
intervention. Answering the target-level question requires examining which consequences
of suppression would persist across interventions and which
depend on the agent or care setting.
Supplement~\ref{supp:stability} illustrates how an explicit
stability assumption can support such an inference.

The CAST example illustrates failure to
resolve causal structures with different
implications for subsequent research.
If ectopy contributes to sudden death but the drugs cause harm
through a separate pathway, another agent might preserve the
benefit while avoiding that harm.
If the harm arises from the mechanism by which these drugs
suppress ectopy, agents sharing that mechanism may share the
risk, directing attention to other means of suppression.
If myocardial damage causes both ectopy and sudden death,
with no causal contribution from ectopy itself, suppression
would provide no survival benefit through that route,
and the target should be reconsidered.
Studies of agents acting through different mechanisms and
investigations of the circumstances and mechanisms of excess
deaths could help distinguish these accounts.
A trial of another agent can constrain the effect of
target modification if its additional effects can be bounded;
Supplement~\ref{supp:bounds} gives an illustration.


\subsection{Delivery, clinical effects, stability and closure}
\label{sec:claims}

For each question below, investigators state what is known,
what they assume and what a further study could clarify.

\textbf{Delivery.} Does the intervention change the intended
target sufficiently and for long enough to produce the proposed
benefit? By phase~III, earlier studies will usually have shown
pharmacological activity. However, showing that a drug reaches
or binds its target does not establish that it produces the
biological change needed for benefit. Tipifarnib illustrates
this distinction: inhibition of farnesyltransferase did not
establish that KRAS function was sufficiently reduced, because
an alternative lipid modification could preserve its activity.
Supplement~\ref{supp:tipifarnib} explains this mechanism and
its implications for pharmacodynamic study design.

\textbf{Clinical effects of target modification.} Does changing
the target affect an outcome that matters to patients?
How large is the effect, and when does it occur?
The \emph{favourable route} describes how the target change
is expected to produce benefit. The same change may also
cause harm through other pathways, which require assessment.

\textbf{Stability.} The regime that generated the evidence
may differ from those in which the intervention will be used.
Which causal relations must remain unchanged to support
the expected clinical effects, and what evidence supports
that assumption? The differences may concern the agent,
population, timing, duration or care.
For example, using evidence from lifelong genetic variation
to predict the effects of treatment begun after disease
develops requires justification. Even when two regimes
modify the same molecular target, the transfer of evidence
must be justified separately for each clinical outcome.

\textbf{Closure.} Could uncertainty about other causal
pathways change the proposed research or treatment decision?
The investigator states which alternative explanations were
considered and whether those still unresolved could change
the proposed action. This judgement may need revision
as new evidence emerges. It does not require every pathway
to be known or the intervention to act through only one.
Failure to detect harm is reassuring only if the studies
would have been likely to detect it had it been large enough
to affect the decision.\cite{mayo2025severity}

The questions are related. A change of setting may alter
how much the intervention changes its target, reduce the
expected benefit or make a harmful pathway clinically
important. Answering these questions may require evidence
from different studies. Figure~\ref{fig:rivals} distinguishes
pathways through the target from other pathways through
which the intervention may affect the outcome; it does not
specify the size of their contributions.

\bigskip

\begin{widefigure}[tbp]
\begin{tikzpicture}[x=1cm,y=1cm]
\foreach \k/\dx/\lab in {1/0/Target route,2/5/Other agent route,3/10/Both routes}{
 \begin{scope}[xshift=\dx cm]
 \node[cnode] (a\k) at (0,0) {$A$};
 \node[cnode] (s\k) at (1.55,0) {$S$};
 \node[cnode] (y\k) at (3.10,0) {$Y$};
 \draw[ar] (a\k)--(s\k);
 \ifnum\k=1\draw[ar] (s\k)--(y\k);\fi
 \ifnum\k>1
  \node[cnode] (z\k) at (1.55,-1.25) {$Z$};
  \draw[ar] (a\k)--(z\k);
  \draw[ar] (z\k)--(y\k);
 \fi
 \ifnum\k=3\draw[ar] (s\k)--(y\k);\fi
 \node[align=center,font=\small] at (1.55,-2.10) {\lab};
 \end{scope}}
\end{tikzpicture}
\caption{Reduced causal structures relevant to a treatment contrast. $A$ is assignment, $S$ the target profile, $Y$ the clinical outcome and $Z$ other agent-related processes. Arrows permit effects of either sign. A favourable total effect can arise through different routes; a small total effect can conceal opposing contributions. Baseline common causes and regimes are omitted here, so these schematics alone do not provide a mediation identification argument.}
\label{fig:rivals}
\end{widefigure}

\bigskip

\subsection{What a causal graph asserts}
\label{sec:graph-interpretation}

Under the graphical semantics used here, 
the graph represents a family of distributions
over the represented set of domain
variables. This family is defined by
the conditional independence relationships dictated by the missing edges
in the graph. For example,
the graph $A\to S\to Y$, with no additional edges or latent
common causes, specifies the factorisation
$p(a,s,y)=p(a)p(s\mid a)p(y\mid s)$ and hence
the conditional independence relationship
$Y\indep A\mid S$.\cite{dawid2021}
A causal interpretation of
the graph additionally requires assumptions
about how interventions alter the mechanisms represented in it. Under one common interpretation, interventions
are modular: they change the mechanisms they target while
leaving the conditional distribution of each other variable
given its parents unchanged. Applied to the above example, modularity means that an intervention setting $A$ to $a$ replaces $p(a)$ by a point mass at $a$ and leaves $p(s\mid a)$ and $p(y\mid s)$ unchanged. If the intervention also sets $S$ to $s$, the distribution of $Y$ becomes $p(y\mid s)$, the same for every value of $a$: at fixed $S$, $A$ has no effect on $Y$. This is the exclusion asserted by omitting the $A\to Y$ arrow. With the arrow, the factor for $Y$ would be $p(y\mid a,s)$, and setting $A$ could change $Y$ at fixed $S$. 
A mere judgement that effects bypassing $S$ are clinically negligible does not justify such an exact restriction. The arrow can instead be retained, with the accompanying prose specifying the evidence and bounds on its effect relevant to the decision.

We let the term \emph{regime} specify the conditions under which a set of
variables are observed or generated. Regimes may differ in how treatment
is assigned, e.g., randomized or external intervention vs routine care. Or they may differ in the intervention
itself or in the surrounding care setting. Or they may differ by the geographical location. These are just a few examples. A \emph{regime indicator} labels these conditions and, incorporated in the causal graph, it points to relations that may differ across
regimes. More
precisely, an arrow from a regime indicator 
node to a variable allows the variable's conditional distribution given its parents in the graph, which we call its \emph{kernel}, to differ across the specified regimes. The absence of such an arrow asserts that the kernel is the same in every specified regime,
which may be adopted as an approximation when plausible 
differences are judged too small to affect the clinical decision. The
accompanying prose should state when this approximation is being
made and explain why those differences can reasonably be disregarded.

A graph shows which effects may be present and leaves their magnitude and timing unspecified. Where competing causal explanations differ
in these quantitative aspects, the accompanying prose must state the magnitude and timing each causal structure assumes, the predictions that follow, and the measurements needed to distinguish the structures.

A graph used in this way supports the two kinds of reasoning distinguished in the Introduction. Identification theory is well developed,\cite{pearl1995,dawidDidelez2010} and in this article it appears explicitly only in Supplement~\ref{supp:stability}, where a stability assumption identifies the effect of a target intervention. The article makes more use of plausible reasoning, which can distinguish causal explanations even where the corresponding effects are not identified; Section~\ref{sec:plausible} develops this relatively unexplored use of graphs.

\subsection{Causal graphs and plausible reasoning}
\label{sec:plausible}

P\'olya's basic pattern of plausible reasoning begins with a deductive step: a conjecture implies a certain consequence. Observing that consequence can then make the conjecture more credible.\cite{polya1945} In our setting the conjecture is a causal account, represented by a graph that states assumptions about causal pathways and by a quantitative model. Deriving the account's predictions, i.e., observable consequences, is P\'olya's deductive step. An observation supports one account over another if it is more expected under the former, given the evidence already available. An observation already implied by that evidence adds no support to any of the accounts. An observation can also favour one account while contradicting both, in which case the accounts compared should be revised (Section~\ref{sec:assessment}). Structural uncertainty is represented by competing graphs, which differ in the presence or absence of causal pathways. The modeller elaborates these graphs, adding variables and pathways where needed, until, where possible, those competing accounts that are material to the decision at hand predict different results for some feasible intervention or observation, so that they can be ranked by plausibility. We illustrate four uses of a graph in this reasoning through Figure~\ref{fig:plausible}, which anticipates the lipid-lowering example
of Section~\ref{sec:complementary}.

In this example two perturbations, $X_1$ and $X_2$, act on different targets, respectively
$T_1$ and $T_2$. The two effects converge on a shared variable $K$,
which affects a measured intermediate $M$; $M$ in turn affects
an outcome $Y$. 
Let "the conjecture" be that the
effects of both
perturbations on $Y$ are
completely mediated
by $M$, as represented by
removing the brown edges
$T_1\to Y$, $T_2\to Y$
and $K\to Y$ from
Figure~\ref{fig:plausible}. The rival conjecture corresponds to the full graph,
brown edges included. For this
illustration, assume that
under "the conjecture" the conditional
mean of $Y$ given $M$ is
$a+bM$, with the same unknown coefficients
$a$ and $b$, whether $X_1$, $X_2$ or
neither is applied, in the same population and with background care, exposure duration and outcome follow-up held fixed.

\emph{Verification of a consequence.}
Under "the conjecture", by
virtue of the above assumptions, a perturbation that changes the mean
of $M$ by $\Delta M$ changes the mean of $Y$ by $b\Delta M$.
For nonzero changes in $M$, under
the conjecture, the effects on $Y$ per unit change
in $M$ must therefore be equal for $X_1$ and $X_2$.
According to the previously discussed
plausible reasoning pattern, an observed agreement between the estimated effects of $X_1$ and $X_2$
on $Y$ per unit change in $M$ supports "the conjecture"
relative to rivals that predict different values, provided
the estimates are precise enough to distinguish those
predictions. A discrepancy beyond that expected from sampling
uncertainty challenges "the conjecture" and directs attention
to its causal exclusions, functional form or stability.

\emph{Verification of an improbable consequence.}
Agreement between the effects
of $X_1$ and $X_2$ on $Y$
per unit change in $M$ carries greater
weight when it
is expected under "the conjecture" but
unlikely under a specified
rival, given the evidence already available.
For example,
suppose the brown-coloured, target-specific,
routes $T_1\to Y$ and $T_2\to Y$
contribute additional effects alongside a common contribution
through $M$. Such agreement then requires these additional contributions
per unit change in $M$ to be similar, which
might be judged unlikely in the specific context.
Agreement weighs against this rival when substantive and
quantitative considerations make that similarity unlikely.

The shared route $K\to Y$ offers a different explanation of a possible agreement
between the estimated effects of $X_1$ and $X_2$
on $Y$ per unit change in $M$. If changes in $K$ produce proportional changes in
$M$ and $Y$, both the $X_1$ and $X_2$
perturbations can have the same effect on $Y$
per unit change in $M$, even if $M$ has no causal effect on $Y$.
The agreement alone therefore does not distinguish this account
from complete mediation through $M$.

\emph{Inference from analogy.}
P\'olya's account of analogy directs attention to relations
shared by otherwise different objects.\cite{polya1945} Here is an example
of analogy. Suppose a genetic variant
reduces the activity of target
$T$ and thereby lowers $M$. If a drug inhibits the same activity in the same tissues,
the genetic finding gives us reason to expect the drug
to change $M$ in the same direction.
The strength of this expectation
depends on whether the
mechanism linking $T$ to $M$ is also
likely to operate
when inhibition begins later in life and lasts for a
shorter period. If $M$ is itself a clinical outcome for which lower
values are preferable, the analogy supports an
expectation of clinical benefit. How much benefit to
expect depends on the degree of target
inhibition, when
it begins and how long it lasts.

\emph{Choosing a study to distinguish the causal structures.}
A useful next study should examine the
effects of an intervention for which
the competing explanations predict different outcomes.
With reference to
our example, another perturbation through $K$
may reproduce the same
effect on $Y$ per unit change in $M$ under each of the following structures:
$M$ causes the change in $Y$, or $K$ changes $M$ and $Y$ through
separate routes. To distinguish these 
structures, a study
could instead change $M$ while leaving $K$ unchanged and
excluding other routes from the intervention to $Y$,
as the perturbation $X_3$ in the figure does.
If the conditional mean relation stipulated
above also holds under $X_3$, complete mediation through $M$
predicts a change in the mean of $Y$ equal to $b\Delta M$.
The rival in which $K$ affects $Y$ but $M$ has no causal
effect predicts no change in $Y$. A study of $X_3$ can thus
distinguish accounts that agree on the predictions for
$X_1$ and $X_2$.

\bigskip

\begin{widefigure}[tbp]
\begin{tikzpicture}[x=1cm,y=.85cm]
\node[cnode] (x1) at (0,1.5) {$X_1$};
\node[cnode] (x2) at (0,-1.5) {$X_2$};
\node[cnode] (t1) at (2.3,1.5) {$T_1$};
\node[cnode] (t2) at (2.3,-1.5) {$T_2$};
\node[cnode] (k) at (4.6,0) {$K$};
\node[cnode] (m) at (6.9,-1.2) {$M$};
\node[cnode] (y) at (9.2,0) {$Y$};
\node[lnode] (x3) at (6.9,-2.5) {$X_3$};

\draw[ar] (x1)--(t1);
\draw[ar] (x2)--(t2);
\draw[ar] (t1)--(k);
\draw[ar] (t2)--(k);
\draw[ar] (k)--(m);
\draw[ar] (m)--(y);

\draw[ar,draw=brown!85!black] (k)--(y);
\draw[ar,draw=brown!85!black]
  (t1.north east) to[out=20,in=120] (y.north);
\draw[ar,draw=brown!85!black]
  (t2.south east)
  .. controls (4.4,-4.3) and (9.2,-4.3) ..
  (y.south);

\draw[bd] (x3)--(m);
\path (-0.5,-3.9)--(9.8,3.0);
\end{tikzpicture}

\caption{A stylized graph for plausible reasoning.
Perturbations $X_1$ and $X_2$ act on targets $T_1$ and $T_2$,
whose effects converge on $K$ and proceed through $M$ to $Y$.
The perturbations are compared in the same population under
common care and timing conditions. Brown arrows mark possible
routes bypassing $M$: $T_1\to Y$ and $T_2\to Y$ branch before
the paths converge, whereas $K\to Y$ branches after.
Removing all brown arrows gives the complete-mediation
conjecture. Retaining $K\to Y$ while removing $M\to Y$, $T_1\to Y$
and $T_2\to Y$ gives the rival in
which both $M$ and $Y$
respond to $K$, but $M$ has no causal effect on $Y$.
The dashed node and arrow mark a proposed additional
perturbation $X_3$, which changes $M$ while leaving $K$
unchanged and has no other effect on $Y$.
Transfer between genetic and pharmacological settings
requires additional assumptions discussed in the text.}
\label{fig:plausible}
\end{widefigure}

\bigskip

\section{From trial effects to causal pathways: the contribution of observational evidence}
\label{sec:complementary}

The statin and PCSK9-inhibitor trials estimated the effect of assigning each drug on specified clinical outcomes, without establishing which biological pathways mediate that effect. \cite{fourS1994,sabatine2017fourier}
This uncertainty may not prevent adoption of a drug, but it may matter when research continues on another agent or seeks to preserve benefit while reducing harm. Lipid-lowering drugs may affect several biological processes, and we ask which of these contribute to fewer cardiovascular events and which contribute to an increased risk of diabetes. The answers could guide research on treatments that
retain the cardiovascular benefit with less risk of diabetes.
Throughout, $Y_C$ denotes cardiovascular events and $Y_D$ denotes diabetes.

\subsection{Making the competing pathways visible}
\label{sec:genetic-cardiovascular}

Causal graphs make explicit the reasoning by which the studies discussed in Section~\ref{sec:statins} bear on questions highlighted in the preceding section.  We show this through a reconstruction of the statin and PCSK9 evidence.

The causal graphs of Figure~\ref{fig:statins-rivals} show both molecular targets explicitly. $H$
denotes the relevant HMGCR activity profile and $P$ the relevant PCSK9
activity profile. The drugs $A_H,A_P$ and the corresponding genetic perturbations
enter at these different nodes. $R$ is LDL-receptor availability, through which
both targets lower LDL, and the arrow $H\to P$ represents the rise in PCSK9
that HMGCR inhibition induces. $L$ is the LDL exposure profile and $W$ the
bodyweight or adiposity profile. The regime indicator $\sigma$ distinguishes
the genetic from the pharmacological regime. Supplement~\ref{supp:genetic} discusses the latent background and the validity of the genetic instruments. The two panels concern
cardiovascular events
and diabetes respectively; their outcome symbols are never interchangeable.

Drug-target Mendelian randomisation already uses genetic proxies of drug targets to interpret trial benefits and harms, as Swerdlow and colleagues did for statins and diabetes.\cite{swerdlow2015} Our addition is to carry this interpretive use of genetic proxies into the choice of the next discriminating study.

\bigskip

\tikzset{
 bhnode/.style={draw=noderule,fill=nodefill,rounded corners=2pt,
minimum width=8mm,minimum height=7mm,inner sep=3pt,
font=\sffamily\normalsize,align=center},
 bhtarget/.style={bhnode,minimum width=15mm,font=\sffamily\small},
 bhhidden/.style={bhnode,densely dotted,fill=white},
 bhreg/.style={bhnode,densely dashed,fill=white},
 bhu/.style={draw=gray,densely dotted,circle,fill=white,inner sep=2pt,font=\sffamily\small,text=gray!40!black},
 bha/.style={-{Latex[length=1.8mm]},line width=.65pt},
 bhua/.style={-{Latex[length=1.5mm]},line width=.5pt,draw=gray},
 bhby/.style={bha,draw=brown!85!black},
 bhsig/.style={bha,densely dashed},
 bhriv/.style={bha,densely dashed,draw=red!65!black},
 bhlab/.style={font=\sffamily\scriptsize\itshape,text=red!65!black,fill=white,inner sep=1pt}}

\begin{widefigure}[p]
\foreach \k/\panel/\heading/\outcome in {0/a/{Cardiovascular events}/{Y_C},1/b/{New-onset diabetes}/{Y_D}}{
\begin{minipage}{\linewidth}
\centering
\textbf{(\panel) \heading}\quad{\small Within baseline stratum $C$}\\[2mm]
\begin{tikzpicture}[x=1cm,y=.70cm]
\node[bhreg] (sig) at (-2.6,0) {$\sigma$};
\node[bhnode] (ah) at (0,1.25) {$A_H$\\[-1pt]{\scriptsize statin}};
\node[bhnode] (ap) at (0,-1.25) {$A_P$\\[-1pt]{\scriptsize evolocumab}};
\node[bhhidden] (ghs) at (0,3) {$G_H^{*}$};
\node[bhnode] (gh) at (-2.6,3) {$G_H$};
\node[bhhidden] (gps) at (0,-3) {$G_P^{*}$};
\node[bhnode] (gp) at (-2.6,-3) {$G_P$};
\node[bhtarget] (h) at (2.8,2.1) {$H$\\HMGCR};
\node[bhtarget] (p) at (2.8,-2.1) {$P$\\PCSK9};
\node[bhnode] (zh) at (2.8,4.3) {$Z_H$};
\node[bhnode] (zp) at (2.8,-4.3) {$Z_P$};
\node[bhnode] (r) at (5.0,0) {$R$\\[-1pt]{\scriptsize receptor}};
\node[bhnode] (l) at (7.3,1.6) {$L$\\[-1pt]{\scriptsize LDL}};
\node[bhnode] (y) at (9.7,0) {$\outcome$};
\draw[bhsig] (sig)--(ah); \draw[bhsig] (sig)--(ap);
\draw[bhsig] (sig.west) -- (-3.6,0) -- (-3.6,-5.5) -- (11.1,-5.5) -- (11.1,0) -- (y.east);
\draw[bha] (ghs)--(gh); \draw[bha] (ghs)--(h);
\draw[bha] (gps)--(gp); \draw[bha] (gps)--(p);
\draw[bha] (ah)--(h); \draw[bha] (ah)--(zh);
\draw[bha] (ap)--(p); \draw[bha] (ap)--(zp);
\draw[bha] (h)--(p);
\draw[bha] (h.east) -- (r); \draw[bha] (p.east) -- (r); \draw[bha] (r)--(l); \draw[bha] (l)--(y);
\draw[bha] (zh.east) to[out=0,in=75] (y.north east);
\draw[bha] (zp.east) to[out=0,in=-90] (y.south);
\draw[bhby] (h.north east) to[out=20,in=105] (y.north);
\draw[bhby] (r) -- (y);
\draw[bhby] (p.south east) to[out=-20,in=-120] (y.south west);
\ifnum\k=1
\node[bhnode] (w) at (7.3,-1.6) {$W$\\[-1pt]{\scriptsize adiposity}};
\draw[bha] (h.east)
  to[out=0,in=90,looseness=1.2]
  ([xshift=2mm]w.north west);
\draw[bha] (p)--(w);
\draw[bha] (w)--(l); \draw[bha] (w)--(y);
\fi
\path (-3.8,-5.7)--(11.3,5.0);
\end{tikzpicture}
\end{minipage}\par\vspace{2mm}}
\caption{Molecular targets, drug actions and two distinct clinical outcomes.
$A_H,A_P$ denote assignments to the statin and PCSK9-inhibitor interventions
relative to their comparators; they need not be randomised jointly. The regime
indicator $\sigma$ distinguishes the genetic regime, in which no drug is assigned,
from the trial regime, in which assignment is randomised. Its dashed arrows permit
regime-dependent kernels for assignment and for the outcome, so that genetic odds
ratios are not transported numerically; its missing arrows into $H,P,R,L$ and $W$
assert that these kernels are shared by the two regimes.
These are working stability assumptions, not consequences
of acting on the same molecular target; their adequacy
depends on how fully the profiles encode tissue, timing
and duration.
$H,P$ denote activity profiles, not genotypes.
The symbol $R$ denotes LDL-receptor availability, through which both
targets lower LDL; $L,W$ are LDL and adiposity profiles. The arrow $H\to P$ represents
the rise in PCSK9 induced by HMGCR inhibition, and $W\to L$ permits an effect of
adiposity on LDL. $Z_H,Z_P$ collect other drug-related processes. Brown arrows permit
routes from the targets and from $R$ that bypass the displayed intermediates; they do
not assert immediate molecular effects. In particular, $P\to W$ is an unresolved
possibility. Functional variation $G_j^{*}$ informs the observed marker or score $G_j$
through a reduced proxy representation. A single latent vector $U$, drawn in
Figure~\ref{fig:genetic-rivals}, is a parent of every node in both panels other than
$\sigma$, the assignments and the genetic variables. Both panels are drawn within a
baseline stratum $C$ that includes ancestry. The arrows form a set of candidate routes;
a missing arrow asserts the absence of an effect. Submodels delete specified edges as
discussed in the text.}
\label{fig:statins-rivals}
\end{widefigure}

\bigskip

\subsection{Using genetic evidence to interpret trial results}
\label{sec:statins}

The Scandinavian Simvastatin Survival Study (4S) randomised
4,444 patients with coronary heart disease to placebo or
simvastatin---a drug that lowers low-density lipoprotein
cholesterol (LDL) by inhibiting HMGCR, an enzyme involved
in cholesterol synthesis. Over a median follow-up of
5.4 years, major coronary events occurred in 431 patients
assigned simvastatin and 622 assigned placebo
(relative risk 0.66; 95\% CI 0.59--0.75).\cite{fourS1994}
This establishes a coronary benefit of assignment to
simvastatin in the population studied. In panel (a) of
Figure~\ref{fig:statins-rivals}, the trial estimates the
total effect of $A_H$ on $Y_C$, summed over the route
through $L$, the brown edges and the drug-related route
through $Z_H$. Represented in panel (a) is also
PCSK9, a protein that
promotes the degradation of hepatic LDL receptors.
Inhibiting PCSK9 preserves these receptors, allowing
the liver to remove more LDL cholesterol from the blood.
HMGCR and PCSK9 therefore provide different biological
targets for lowering LDL, and in the graph both act on
$L$ through $R$, the availability of LDL receptors.
Here $R$ includes receptor availability in the liver
and in other relevant tissues. Comparable changes
in hepatic receptor availability need not imply
comparable changes in pancreatic beta cells.

Panel (b) of the same figure represents the corresponding pathways to
new-onset diabetes, $Y_D$, and additionally includes
adiposity, $W$. Perturbing either target, HMGCR or PCSK9, may affect
diabetes risk through LDL, through routes that bypass
LDL, or through changes in adiposity, which may themselves
affect LDL. The arrow $P\to W$ represents an unresolved
possibility.

The HMGCR and PCSK9 genetic scores combine variants
used as proxies for inhibition of the respective
targets.\cite{burgess2013scores}
We assume that each score is associated with the intended
target perturbation, is independent of confounders of
the target--outcome relation, and affects the outcomes
only through that target. These are the relevance,
independence and exclusion restriction assumptions,
respectively. The graph represents the proposed entry
of each genetic perturbation at its target; the rival
structures in Figure~\ref{fig:genetic-rivals} illustrate
possible violations of these assumptions.

Consider first the hypothesis of total LDL mediation
of the coronary effect (TLMCE), asserting that both genetic perturbations
(of HMGCR and PCSK9) reduce coronary risk entirely through
LDL lowering. In panel (a), TLMCE is the submodel without
the brown edges, in which every path from $H$ or $P$ to
$Y_C$ passes through $L$. To obtain a prediction on the scale of the reported associations,
add the quantitative working assumption that comparable changes
in lifelong LDL exposure produce approximately proportional
changes in coronary log odds, with the same proportionality
for both scores in comparable populations.
Together with TLMCE and the instrument assumptions, this
proportionality assumption predicts approximately equal coronary
associations for the HMGCR and PCSK9 scores per unit of
genetically predicted LDL reduction (P\'olya's deductive step). This concordance
has been corroborated by genetic evidence, obtained by Ference and colleagues
from a study of 112,772 participants. For each 10 mg/dL lower genetically predicted LDL
cholesterol, the estimated odds ratio for myocardial
infarction or coronary death was 0.81 for both scores
(95\% CI 0.72--0.90 for HMGCR and 0.74--0.89 for
PCSK9).\cite{ference2016pcsk9}
The concordance supports the joint account given by
TLMCE, the instrument assumptions and the common
proportionality assumption, relative to alternatives
that predict unequal scaled associations. It favours this account over
target-specific bypass explanations to the extent that those
explanations predict different coronary associations per unit
of LDL reduction. Under appropriate scaling,
concordance could also be produced by similar bypass contributions.
The plausibility of such a scenario must be assessed
from biological and quantitative evidence.
This concordance does not, however, distinguish LDL
mediation from an effect through $R\to Y_C$:
both targets lower LDL through $R$.
If comparable LDL reductions reflect comparable changes
in receptor availability, a shared effect through
$R\to Y_C$ could also produce the observed concordance between the two scores.
The comparison of the HMGCR and PCSK9 scores therefore cannot
distinguish between TMLCE and this last explanation. A perturbation that
lowers LDL while leaving receptor availability $R$ unchanged could. This
is the comparison described in
Section~\ref{sec:plausible}: HMGCR and PCSK9 correspond
to $T_1$ and $T_2$, receptor availability to $K$,
and LDL to $M$.

The comparison can be extended beyond HMGCR and PCSK9.
Silverman and colleagues found similar reductions
in major vascular risk per unit of LDL lowering
with statins and established nonstatin interventions
acting primarily through LDL-receptor
upregulation.\cite{silverman2016}
Genetic comparisons across HMGCR, NPC1L1, PCSK9
and LDLR show similar concordance.\cite{ference2017eas}
These findings strengthen the case against
target-specific explanations that predict unequal
effects per unit of LDL lowering. However, because
the perturbations share the receptor pathway, they
do not distinguish LDL mediation from a shared
effect through $R\to Y_C$.

The FOURIER trial established the clinical benefit
of adding evolocumab to statin therapy. Evolocumab
is an antibody that lowers LDL by inhibiting PCSK9.
Among 27,564 patients with atherosclerotic cardiovascular
disease, the primary cardiovascular composite occurred in
9.8\% assigned evolocumab and 11.3\% assigned placebo over
a median 2.2 years (hazard ratio 0.85; 95\% CI
0.79--0.92).\cite{sabatine2017fourier}
In the graph, FOURIER perturbs $P$ through $A_P$ in the
trial regime, on a background of statin therapy acting
through $A_H$. Under TLMCE, further LDL lowering through PCSK9 inhibition
would be expected to reduce coronary events, provided that
the LDL-mediated benefit also operates under drug treatment
over the exposure duration considered.
\cite{ference2016pcsk9,ference2012,ference2017eas}
The cardiovascular benefit observed in FOURIER is consistent
with this expectation. It further supports TLMCE over
HMGCR-specific explanations of statin benefit to the extent
that those explanations predict less benefit from PCSK9
inhibition for a comparable magnitude and duration of
LDL reduction.

Swerdlow and colleagues
combined evidence from 43 genetic studies and 20
statin trials. In the trials, involving 129,170
participants, statin initiation or intensification
increased new-onset diabetes (odds ratio 1.12;
95\% CI 1.06--1.18). HMGCR variants associated with
lower LDL also showed associations with higher
bodyweight and diabetes risk.\cite{swerdlow2015}
In panel (b), the trials estimate the total effect of $A_H$
on $Y_D$, which may pass through the target $H$ or
through the drug-specific processes $Z_H$. The genetic perturbations involve no drug and therefore
do not act through  $Z_H$.
Under the assumptions that the variants affect diabetes
through $H$ and that the mechanism carries over to drug treatment, the concordant
findings support a contribution of HMGCR inhibition
to the adverse effect of statins. Figure~\ref{fig:diabetes-explanations} presents four
explanations of the excess diabetes risk. E1 attributes
it to LDL lowering itself. E2 attributes it to increased
cholesterol uptake through greater LDL-receptor
availability. E3 attributes it to weight gain caused
by both target perturbations. E4 attributes it to an
HMGCR-specific effect that bypasses receptor availability,
LDL and weight. These mechanisms can coexist.
Table~\ref{tab:diabetes-rivals} gives the predictions
when each mechanism acts alone.

We compare the diabetes odds ratios for the HMGCR
and PCSK9 scores, scaled to the same reduction in
genetically predicted LDL. Under E1, we expect similar
odds ratios if the LDL changes have comparable effects
on diabetes in the populations studied. E2 predicts
similar odds ratios if equal LDL reductions correspond
to comparable receptor changes and those changes have
comparable effects on diabetes. E3 predicts similar
odds ratios if the weight-mediated effects are comparable
at that scaling. E4 alone predicts an excess for HMGCR
and none for PCSK9. An HMGCR-specific contribution
in addition to a shared mechanism would predict a
higher odds ratio for HMGCR if the shared contributions
are comparable and no other effects offset the difference.

Ference and colleagues reported diabetes odds ratios
of 1.13 (95\% CI 1.06--1.20) for HMGCR and 1.11
(95\% CI 1.04--1.19) for PCSK9 per 10 mg/dL lower
genetically predicted LDL.\cite{ference2016pcsk9}
The excess associated with PCSK9 is difficult to
reconcile with E4 as the sole explanation. The similarity
of the estimates does not distinguish E1 from E2 or
E3 and does not rule out a smaller additional
HMGCR-specific contribution.

Ference and colleagues suggested a shared LDL-receptor
mechanism, supported also by their analysis of LDLR
variants.\cite{ference2016pcsk9} Besseling and colleagues
found a lower prevalence of diabetes in familial
hypercholesterolaemia and proposed that reduced
receptor-mediated cholesterol uptake could protect
pancreatic beta-cell function.\cite{besseling2015}
This observation is consistent with E2, but does not
distinguish it from E1.

Two uncertainties remain. The first is whether PCSK9
perturbation increases bodyweight, as the shared
weight-mediated explanation E3 requires. The evidence
on this relation remains inconclusive
(Supplement~\ref{supp:genetic}).

The second concerns whether the diabetes risk associated
with lifelong genetic perturbation also occurs during
shorter drug treatment. In the prespecified FOURIER
diabetes analysis, the hazard ratio for new-onset diabetes
with evolocumab was 1.05 (95\% CI 0.94--1.17).
\cite{sabatine2017diabetes}
The trial therefore did not establish an increased risk,
but its estimate remains compatible with a modest increase.

Figure~\ref{fig:entry-points} shows where the comparisons
in Table~\ref{tab:diabetes-rivals} enter the graph.
A perturbation that lowers LDL while leaving receptor
availability unchanged could distinguish E1 from E2,
provided that it has no other pathway to diabetes.
The corresponding coronary comparison would ask
whether lowering LDL reduces coronary risk when
receptor availability is unchanged.

LDLR loss-of-function variants change both receptor
function and LDL exposure. The comparison therefore
does not by itself distinguish E1 from E2. Preventing
weight gain during target perturbation could test E3.
If weight gain accounts for the entire excess diabetes
risk, fully preventing that gain should remove the
excess, provided that the intervention has no effect
on diabetes outside the pathways through $W$. Any
effect of weight gain through subsequent changes in
LDL is part of the weight-mediated effect.

The causal record below selects a genetic comparison
that lowers LDL while leaving receptor availability
unchanged, to help distinguish E1 from E2.

\bigskip

\begin{widefigure}[tbp]
\scalebox{0.85}{%
\begin{tikzpicture}[x=1cm,y=.8cm]
\node[bhnode] (gh) at (0,1.8) {$G_H$};
\node[bhnode] (gp) at (0,-1.8) {$G_P$};
\node[bhtarget] (h) at (2.6,1.8) {$H$\\HMGCR};
\node[bhtarget] (p) at (2.6,-1.8) {$P$\\PCSK9};
\node[bhnode] (r) at (5.4,0) {$R$\\[-1pt]{\scriptsize receptor}};
\node[bhnode] (l) at (8.2,-1.3) {$L$\\[-1pt]{\scriptsize LDL}};
\node[bhnode] (w) at (6.2,3.0) {$W$\\[-1pt]{\scriptsize adiposity}};
\node[bhnode] (y) at (11.0,0) {$Y_D$};
\node[bhnode,fill=white] (fh) at (5.4,-2.3) {LDLR\\[-1pt]{\scriptsize loss of function}};
\node[bhreg] (al) at (8.2,-3.3) {$A_L$\\[-1pt]{\scriptsize receptor-independent}};
\node[bhreg] (aw) at (6.2,4.7) {$A_W$\\[-1pt]{\scriptsize weight prevented}};
\draw[bha] (gh)--(h); \draw[bha] (gp)--(p);
\draw[bha] (h)--(p);
\draw[bha] (h)--(r); \draw[bha] (p)--(r);
\draw[bha] (r)--(l); \draw[bha] (l)--(y.south west);
\draw[bha] (h.north east)--(w.west);
\draw[bha,densely dashed] (p.north east)--(w.south west);
\draw[bha] (w)--(y.north east);
\draw[bhby] (r.east)--(y.west);
\draw[bhby] (h.east) -- (y.north);
\draw[bha] (fh)--(r);
\draw[bhsig] (al)--(l);
\draw[bhsig] (aw)--(w);
\path (-0.5,-3.9)--(11.6,5.2);
\end{tikzpicture}
}
\caption{Entry points for the comparisons in
Table~\ref{tab:diabetes-rivals}, shown on a simplified
version of panel (b) of Figure~\ref{fig:statins-rivals}.
The genetic scores enter at $H$ and $P$, and LDLR
loss-of-function variants enter at $R$. The proposed
interventions are shown with dashed arrows. $A_L$ lowers
LDL while holding receptor availability $R$ unchanged,
and $A_W$ prevents weight gain. The brown edges mark
the receptor pathway $R\to Y_D$ and the HMGCR-specific
pathway $H\to Y_D$. The dashed edge $P\to W$ indicates
uncertainty about whether PCSK9 perturbation affects
weight. The LDL-lowering pathways from $H$ and $P$
converge at $R$, so comparing the two scores can help
assess contributions from pathways that diverge before
that point. LDLR loss of function changes both receptor
function and LDL exposure, so this comparison cannot
by itself distinguish the receptor pathway from the
pathway through LDL. Drug nodes, the regime indicator,
and the edges $W\to L$ and $P\to Y_D$ are omitted for clarity.}
\label{fig:entry-points}
\end{widefigure}

\bigskip

\begin{widefigure}[tbp]
\centering
\begin{minipage}[t]{.48\linewidth}
\centering\textbf{E1. LDL lowering}\par\medskip
\begin{tikzpicture}[x=.95cm,y=1cm]
\exnodes{exmuted}
\draw[exar] (h)--(r); \draw[exar] (p)--(r);
\draw[exar] (r)--(l); \draw[exar] (l)--(y);
\path (-.5,-1.7) rectangle (6.3,2);
\end{tikzpicture}
\end{minipage}\hfill
\begin{minipage}[t]{.48\linewidth}
\centering\textbf{E2. Receptor-mediated uptake}\par\medskip
\begin{tikzpicture}[x=.95cm,y=1cm]
\exnodes{exmuted}
\draw[exar] (h)--(r); \draw[exar] (p)--(r);
\draw[exback] (r)--(l);
\draw[exar] (r.north east)--(y.west);
\path (-.5,-1.7) rectangle (6.3,2);
\end{tikzpicture}
\end{minipage}
\par\vspace{12mm}
\begin{minipage}[t]{.48\linewidth}
\centering\textbf{E3. Weight gain: shared mechanism}\par\medskip
\begin{tikzpicture}[x=.95cm,y=1cm]
\exnodes{exnode}
\draw[exar] (h)--(w);
\draw[exar] (p.north east)--(w.south west);
\draw[exar] (w)--(y);
\draw[exback] (r)--(l);
\path (-.5,-1.7) rectangle (6.3,2);
\end{tikzpicture}
\end{minipage}\hfill
\begin{minipage}[t]{.48\linewidth}
\centering\textbf{E4. HMGCR-specific effect}\par\medskip
\begin{tikzpicture}[x=.95cm,y=1cm]
\exnodes{exmuted}
\draw[exback] (h)--(r); \draw[exback] (p)--(r);
\draw[exback] (r)--(l);
\draw[exar] (h.south east) .. controls (2,.5) and (4,.5) .. (y.west);
\path (-.5,-1.7) rectangle (6.3,2);
\end{tikzpicture}
\end{minipage}
\par\medskip
{\small\textcolor{route}{\rule{12mm}{1.1pt}}\quad Proposed pathway to diabetes
\qquad\textcolor{muted}{\rule{12mm}{.55pt}}\quad LDL pathway shown for context}\par
\captionsetup{font=small,labelfont=bf}
\caption{Four proposed explanations of the excess diabetes risk associated with target perturbation. $H$: HMGCR activity; $P$: PCSK9 activity; $R$: LDL-receptor availability; $L$: LDL exposure; $W$: bodyweight or adiposity; $Y_D$: diabetes. Each panel shows the pathway proposed to carry the excess risk, with nodes in fixed positions. In E1, both targets act through $R$ and $L$. In E2, the effect of $R$ on diabetes bypasses circulating LDL. E3 shows the shared weight-mediated explanation: both target perturbations increase weight, which increases diabetes risk. If only HMGCR perturbation increased weight, weight mediation alone could explain an excess for HMGCR but not for PCSK9. E4 represents an HMGCR-specific effect bypassing $R$, $L$ and $W$, and cannot alone explain the PCSK9 association. These are pathway sketches, not complete causal graphs: $H\to P$, genetic and drug nodes, and other background relations are omitted; the LDL-lowering branches from the targets are also omitted in E3 for clarity. No $W\to L\to Y_D$ pathway is drawn in this simple version of E3; adding it would still give mediation through $W$. Arrows specify causal direction, not the sign or magnitude of effects. Quantitative predictions require the assumptions in Table~\ref{tab:diabetes-rivals}.}
\label{fig:diabetes-explanations}
\end{widefigure}

\bigskip

\clearpage
\begin{widetable}[tbp]
\footnotesize
\setlength{\tabcolsep}{3pt}
\renewcommand{\arraystretch}{1.18}
\caption{Competing explanations of the diabetes associations: predictions for four comparisons and implications for further research. Predictions are conditional on the assumptions stated below.}
\label{tab:diabetes-rivals}
\begin{tabularx}{\linewidth}{@{}L{27mm}L{27mm}L{20mm}L{23mm}L{22mm}Y@{}}
\toprule
Explanation (Fig.~\ref{fig:diabetes-explanations}) & HMGCR versus PCSK9: diabetes ORs per equal LDL reduction & LDLR loss of function & LDL lowering with $R$ unchanged & Target perturbed, weight gain prevented & Implication for further research \\
\midrule
E1. LDL lowering
 & Similar
 & Lower risk
 & Higher risk
 & Excess persists if LDL lowering is maintained
 & Assess the benefit--harm balance of LDL-lowering strategies \\
\addlinespace[6pt]
E2. Receptor-mediated cholesterol uptake
 & Similar
 & Lower risk
 & No change in risk through this mechanism
 & Excess persists if the receptor change is maintained
 & Test LDL lowering that leaves receptor availability unchanged \\
\addlinespace[6pt]
E3. Weight gain caused by both target perturbations
 & Similar if weight-mediated effects are comparable
 & Depends on effects on weight
 & No excess if weight is unchanged
 & Excess removed if weight gain is fully prevented
 & Test whether preventing weight gain prevents the excess diabetes risk \\
\addlinespace[6pt]
E4. HMGCR-specific effect bypassing $R$, $L$ and $W$
 & Excess for HMGCR; none for PCSK9
 & No specific prediction
 & No excess through this mechanism
 & Excess persists for HMGCR
 & Test whether PCSK9 inhibition avoids this HMGCR-specific harm \\
\midrule
Evidence discussed here
 & Similar\cite{ference2016pcsk9}
 & Lower diabetes prevalence in FH\cite{besseling2015}
 & Proposed comparison
 & Proposed comparison
 & \\
\bottomrule
\end{tabularx}
\par\smallskip
\begin{minipage}{\linewidth}
\footnotesize\raggedright
\textit{Assumptions.} Each row treats the stated mechanism as the sole cause of the excess risk under the target perturbation concerned; mixtures remain possible. Predictions for the score comparison require valid instruments, comparable effects of the shared pathways per equal LDL reduction, and no offsetting effects. Under E2, equal LDL reductions must correspond to comparable receptor changes; under E3, they must correspond to comparable weight-mediated effects. Proposed perturbations must have no additional pathways to diabetes; preventing weight gain may also change LDL.\par\smallskip
\textit{Interpretation.} $R$: LDL-receptor availability; OR: odds ratio; FH: familial hypercholesterolaemia. The LDLR comparison concerns loss of receptor function. The cited FH evidence concerns prevalence, which is not itself an estimate of incident risk. E4 alone cannot explain an increased risk associated with PCSK9 variants. A comparison distinguishes explanations only when their predictions differ.
\end{minipage}
\end{widetable}

\bigskip

\subsection{Illustrative causal record}
\label{sec:record-example}

A causal record connects an unresolved causal question to a
research or treatment decision, stating the evidence, assumptions
and findings that would change that decision. We illustrate it
using the lipid-lowering example; Section~\ref{sec:record}
describes its general content and prospective use.

This illustrative design record asks whether the diabetes
associations discussed in Section~\ref{sec:statins} arise from
LDL lowering itself (E1 in Table~\ref{tab:diabetes-rivals}) or
from increased receptor-mediated cholesterol uptake (E2).
The proposed comparison addresses this distinction while
retaining weight mediation (E3) and additional pathways as
alternative explanations.

\medskip

\noindent\textbf{Status and information available.}
This is a retrospective reconstruction prepared by the authors
and adopted by no committee. It uses the cited genetic and trial
evidence available by 31 December 2017. A prospective record
would also state its preparation date and version identifier.

\medskip

\noindent\textbf{Decision.}
We must decide whether to test a drug that lowers LDL
without increasing LDL-receptor availability.
If LDL lowering itself causes the excess diabetes risk
(E1), such a drug would retain that risk.
If increased receptor-mediated cholesterol uptake causes
the excess risk (E2), the drug could avoid it while
preserving the coronary benefit.
The proposed genetic comparison would help assess
this possibility before committing to a trial.
A trial would also require evidence that the candidate
drug is sufficiently safe and that its other effects
would not obscure the comparison between E1 and E2.

\medskip

\noindent\textbf{Causal question.}
Would lowering LDL, while leaving LDL-receptor
availability unchanged, increase the risk of type~2
diabetes and reduce coronary risk? The comparison
of interest is with no LDL lowering in the same
population over the same follow-up period, assuming
that the intervention affects these outcomes only
through LDL. To inform this question, we propose a Mendelian
randomisation comparison between the HMGCR and PCSK9
scores and variants that lower LDL without changing
receptor availability. Outcome associations would
be compared per 10~mg/dL lower genetically predicted
LDL. Associations with bodyweight and liver fat
would help assess alternative pathways.

We do not identify a validated set of variants satisfying
these restrictions here. The record specifies the contrast
to seek; its feasibility depends on finding suitable
instruments.

\medskip

\noindent\textbf{Delivery.}
Each score must have an adequately estimated association with
LDL. For the proposed comparison, receptor-independent lowering
must also leave $R$ unchanged in the relevant tissues.
Independence from receptor-mediated clearance does not by itself
establish this condition. Because receptor availability is
unmeasured in the cohort, the condition requires biological
evidence about the selected variants and their downstream effects.
Only under this restriction can the proposed perturbation be
represented as entering at $L$ in
Figure~\ref{fig:entry-points}.

The contrast treats E2 as an effect of increased receptor
availability that is absent when receptor availability
is unchanged. This is a simplifying assumption:
unchanged receptor availability does not itself guarantee
unchanged cholesterol uptake. If lowering circulating LDL
also changes the uptake relevant to diabetes, the E2
prediction must be revised.

\medskip

\noindent\textbf{Clinical effects and alternatives.}
If LDL lowering itself increases diabetes risk (E1),
variants that lower LDL without changing receptor
availability should also increase that risk.
If the excess risk instead arises from increased
receptor-mediated cholesterol uptake (E2), these
variants should not increase it through that mechanism.

However, an increase in diabetes risk would not
necessarily establish E1. For example, variants that
reduce lipoprotein production might also increase
liver fat, providing another explanation for the
diabetes association. Conversely, no increase in
diabetes risk would be compatible with both E2 and
the weight-mediated explanation E3 if the variants
did not increase weight.

The study would therefore assess associations with
bodyweight and liver fat to help interpret its main
findings. An absence of detectable associations would
not by itself exclude these alternative pathways.

The PCSK9 diabetes association weighs against E4 as the sole
explanation. E3 remains possible, and a null diabetes association
for a perturbation that changes neither receptor availability
nor weight would be compatible with both E2 and E3.

\medskip

\noindent\textbf{Stability.}
To compare the diabetes associations of the genetic
scores per unit of LDL reduction, we assume that the
same reduction over the same period would have the
same LDL-mediated effect on diabetes risk, regardless
of which genetic mechanism produced it.
Otherwise, differences between the scores could
reflect differences in the response to LDL rather
than effects through other pathways.

Transferring the findings from a genetic to a drug
regime requires a further assumption: the mechanism
responsible for the genetic association must also
operate when inhibition begins later in life and
lasts for a shorter period. The genetic findings
could therefore support the choice of a drug to
test, but would not establish the size of its
clinical benefits or harms.

\medskip

\noindent\textbf{Closure.}
Before choosing this study, we must ask whether other
effects of the selected variants could prevent us
from distinguishing E1 from E2. For example, an
increase in diabetes risk could arise from changes
in weight or liver fat, or from effects unrelated
to the intended target. If these explanations
cannot be adequately excluded or their effects
bounded, we should select different variants or
a different study design.

\medskip

\noindent\textbf{Assessment and revision.}
Before analysing the data, we would state what results
would support each explanation. If LDL lowering itself
increases diabetes risk (E1), the proposed genetic score
should show an increase in risk despite leaving receptor
availability unchanged. If increased receptor-mediated
cholesterol uptake causes the risk (E2), the score
should show no increase, provided that it has no other
effect on diabetes.

To determine the required sample size, we would use
diabetes odds ratios of 1.11 under E1 and 1.00 under
E2, per 10~mg/dL lower genetically predicted LDL. The value 1.11 comes from the
PCSK9 comparison.\cite{ference2016pcsk9}
Using it for the proposed score assumes that equal LDL
reductions have comparable effects on diabetes.
The study must be large enough to distinguish these
predictions. If both values were known exactly, a
standard error of about 0.037 for the log odds ratio
would give 80\% power at a two-sided 5\% significance
level. Planning must also allow for uncertainty in
the PCSK9 estimate and in the LDL reduction associated
with each score.

The coronary outcome provides a second comparison.
If LDL lowering accounts for the coronary benefit,
we would expect an odds ratio near 0.81 per 10~mg/dL
lower LDL, assuming an effect comparable to that of
the HMGCR and PCSK9 scores. If the benefit instead
depends entirely on increased receptor availability,
we would expect no coronary benefit when receptor
availability remains unchanged.

We would interpret the findings as follows.
A diabetes odds ratio near 1.11 would support E1 over
E2 if effects through weight, liver fat and other
pathways could be excluded. If diabetes risk and
liver fat both increased, we could not tell whether
LDL lowering or liver fat explained the increased risk.

A diabetes odds ratio near 1.00 would support E2 over
E1. It would also remain compatible with E3 if the
variants did not increase weight. If this result
were accompanied by a coronary odds ratio near 0.81,
it would support testing whether a drug acting through
this mechanism could preserve coronary benefit without
increasing diabetes risk.

These interpretations require sufficiently precise
estimates. A confidence interval that included both
1.00 and 1.11 would leave the diabetes comparison
unresolved at that confidence level. This could occur
even if the study achieved its planned precision.

\bigskip

\section{The causal record in practice}
\label{sec:record}

\medskip

The record in Section~\ref{sec:record-example} shows how
competing explanations of diabetes risk guide the choice
of a genetic comparison. We now describe how investigators
and decision-making bodies can prepare, assess and revise such records.

\subsection{The decision and its evidential grounds}

A causal record specifies the decision and the causal 
comparison needed to inform it, stating the estimand
separately from the assumptions needed to identify
it and the method used to estimate it.\cite{hernan2026estimands}
In a \emph{design record}, the investigators responsible
for the protocol identify the alternatives that could change its
design or interpretation of results. In a \emph{decision record}, a regulator
or guideline panel identifies the evidence and assumptions
supporting the proposed use. Each version names its authors
and adopting body, its date and the information then available.

Table~\ref{tab:minimum} summarises the content, which is the
same in both versions; they differ in the decision recorded (a
research choice in the design record, a treatment or regulatory 
choice in the decision record) and in who adopts the record.
The graph and prose connect evidence from particular
studies to four questions: does the intervention change
its target, what benefits and harms follow, which
effects persist across regimes, and could unresolved
pathways change the decision? Uncertainty about a causal pathway may call for
additional measurements, a study of another
intervention, or observational evidence.
Quantitative forecasts and
model probabilities are optional; the basic record can be written
when numerical probabilities cannot be defended.

\bigskip

\begin{widetable}[tbp]
\small
\caption{What to include in a causal record.}
\label{tab:minimum}
\begin{tabularx}{\linewidth}{@{}L{31mm}Y@{}}
\toprule
Entry & Information to record \\
\midrule

Decision &
State what must be decided, which options are available,
and what benefits, harms and burdens would favour
one option over another. \\

Causal question &
Specify whose outcomes are being compared, under which
interventions or target changes, and over what period.
State whether the question concerns assignment to treatment
or following it. Where relevant, explain how events such
as death affect the outcome being assessed. \\

Causal account &
Use a graph and text to explain how the intervention
changes its target and how that change could produce
benefit or harm. State which effects are assumed to
persist in other settings and whether unresolved
pathways could change the decision. Separate what is
known from what is assumed or remains uncertain. \\

Complementary evidence &
Identify the studies supporting each proposed causal
link. Explain why their findings apply to the question
at hand. Consider differences in measurements, treatment
and patient populations, as well as confounding and
selection. State whether the data cover the patients
and exposures needed for the proposed comparison. \\

Alternatives and design &
Describe other explanations that fit the existing
evidence but would lead to different decisions.
State what further study could distinguish them,
what each explanation predicts, and which explanations
that study would still leave unresolved. \\

Assessment &
Specify what will be measured and when, how precise
the results must be, and what findings would count
against each explanation. Examine whether the conclusion
changes when uncertain assumptions are varied.
If assigning probabilities to explanations, state
the statistical models and priors used to calculate them. \\

Responsibility and revision &
Name the authors and any body adopting the record.
Give its date and the evidence available when it was
written. Retain earlier versions. For each revision,
state what new evidence prompted the change and how
it affects the decision. \\

\bottomrule
\end{tabularx}
\end{widetable}

\bigskip

\subsection{Prospective use, assessment and revision}
\label{sec:assessment}

A causal record may be prepared after one study to guide
the next. Before the new results are known, investigators
state what each explanation predicts, which observations
would challenge it, and whether the study could detect
differences large enough to change the decision. This
requirement accords with Mayo's emphasis on tests capable
of exposing mistaken claims.\cite{mayo2025severity}

A finding that contradicts a prediction may challenge
the proposed pathway, the assumed size or timing of its
effect, or the adequacy of the measurements. For example,
achieving the intended target change without clinical
benefit could mean that the change is ineffective,
that another drug effect offsets its benefit, or that
the benefit takes longer to appear. The next measurements
or comparisons should help distinguish these possibilities.

Unexpected findings may also suggest explanations absent
from the original record. These should be recorded with
the evidence that prompted them. Their ability to explain
those findings does not constitute an independent test:
further evidence should assess predictions that distinguish
them from their rivals. Each revision should state what
has changed and how it affects the next research or
treatment decision.

Quantitative assessment may use confidence regions, predictive
models or other justified procedures. Posterior model probabilities
express relative support within the specified set; even the favoured
account may predict the observations poorly. Gelman and Shalizi
distinguish updating within a model set from checking and revising
that set.\cite{gelmanShalizi2013}

New evidence may change what investigators take the
clinical problem to be, rather than merely alter their
confidence in existing explanations. This possibility
recalls Hegel's account of experience as transforming
the object of inquiry.\cite{hegel1807}.\cite{hegel1807} In patients
with \emph{Helicobacter pylori} infection, evidence concerning
infection and eradication expanded questions about ulcer healing
to include the infectious process and prevention of
recurrence.\cite{hentschel1993} Such a change can alter the target,
population and observations relevant to the next study, as well
as the probabilities assigned to existing explanations.

Another example arose in our analysis of data on subarachnoid haemorrhage.
\cite{berzuini2024sah} The initial question was whether external ventricular
drainage (EVD) would benefit a specified group of patients affected
by acute hydrocephalus. As the investigation developed, the question
shifted: among patients whose aneurysm had not yet been secured,
should EVD precede aneurysm securing or follow it? The intervention
of interest therefore became the timing of EVD relative to aneurysm
securing. This reformulation changed the treatment
strategies to be compared and the evidence needed to assess them.

\bigskip

\section{Relation to existing approaches and evaluation}
\label{sec:evaluation}

\subsection{The proposed contribution}
\label{sec:related}

We propose using causal graphs and an accompanying
record to state which explanations of existing findings
would lead to different research or treatment decisions,
what study could distinguish them, and how its results
would change those decisions. The graphs show the
competing pathways; the text states the assumptions
needed to derive their predictions. The record preserves
this reasoning and explains how new evidence changes it.

Hern\'an et al. compare PICO, the target trial framework
and ICH E9(R1), emphasising the need to specify the causal
estimand before choosing an analysis.
\cite{hernan2026estimands}
Causal discernment builds on that specification.
Several causal accounts may imply the same effect for
one well-defined comparison but different effects for
another agent or treatment strategy. The record states
which further comparison could distinguish these
accounts and how the result would affect the choice
of the next study.

Existing approaches provide much of the basis for this
work. Mechanistic reasoning and EBM+ assess evidence
for mechanisms and its relevance to other populations
or settings.\cite{howick2010,howick2013,parkkinen2018}
Triangulation compares evidence from sources with
different potential biases.\cite{lawlor2016}
Experimental designs for mechanisms and Bayesian
design for model discrimination help select informative
comparisons.\cite{imai2013,hainy2022}
Mendelian randomisation and methods for combining
evidence across regimes provide tools for drawing
causal conclusions from different studies.
\cite{lawlor2008mr,dawidDidelez2010,bareinboim2016}

Our contribution is to organise the use of these
approaches around a decision and to document the
reasoning that connects one study to the next.
We develop the use of graphs in plausible reasoning:
with substantive and quantitative assumptions, they
show why a finding supports one causal account over
another, why some accounts remain indistinguishable,
and what further comparison could separate them.
Regime indicators make explicit which causal relations
are assumed to persist across the settings compared.
The record states how these judgements guide the
research decision and preserves the grounds for
subsequent revisions.

The record could be incorporated into existing
procedures for protocols and recommendations.
SPIRIT 2025 calls for reporting an evidence-supported
scientific rationale, including an explanation of how
the intervention might work.
\cite{spirit2025,spirit2025explanation}
GRADE Evidence-to-Decision frameworks organise the
grounds for recommendations, including effects,
values, resources and feasibility.\cite{grade2016}
Within these procedures, the record would make explicit
which competing causal explanations matter to the
decision and what evidence could distinguish them.

The ability to distinguish explanations is only one
reason to choose a study. Its value also depends on
whether its possible results would change a decision
and improve its consequences enough to justify the
costs and burdens of research.\cite{raiffa1961}
The causal record makes the relevant options,
predictions and assumptions explicit. These can inform
a formal value-of-information analysis when probabilities
and utilities can be justified, or a qualitative
comparison of studies when they cannot.

\subsection{Limitations}

Making assumptions explicit does not establish that
they are correct, and a record may omit an important
causal explanation. It could also become an
administrative exercise without improving decisions.

We have not evaluated the procedure in institutional
use. The examples illustrate its possible uses but
do not establish that it improves research decisions.
Its added value should be assessed against ordinary
rationale review, considering the choice of informative
studies and the time and resources required.
Fewer trials failing their primary endpoint would
not by itself demonstrate improvement: an unfavourable
trial may still resolve an uncertainty important
to the next decision.

\bigskip

\section{Conclusion}

Causal discernment orients clinical research by connecting
competing explanations of existing findings to the choice of subsequent
interventions, measurements and comparisons, while allowing the 
question itself to be revised. Delivery, the clinical effects of target
modification, stability and closure identify what must be investigated 
or justified. Complementary studies can address different parts of
that task: the statin example shows how genetic evidence can inform
target-level questions left open by treatment trials, while
Supplement~\ref{supp:stability} works through one way of
combining trial and observational information.
In the statin example, causal graphs, with substantive and
quantitative assumptions, showed which findings favoured one
explanation over another and where a further intervention
could distinguish them. Regime indicators state which relations
are assumed to persist across the settings being compared. A causal
record preserves these explanations, assumptions and evidential
grounds, making the direction chosen for further research open to 
scrutiny and revision. Its practical value beyond existing review
requires prospective evaluation.

\bigskip

\section*{Data availability}
No participant-level data were generated or analysed. All numerical inputs, causal
assumptions and probability models for the hypothetical calculations are specified 
in the Supporting Information.

\bigskip

\section*{Ethics statement}
This conceptual analysis uses published sources and hypothetical calculations
and involves no recruitment or participant-level data.

\bigskip

\section*{Author contributions}
Carlo Berzuini: conceptualisation (lead), methodology, formal analysis, writing -- original
draft. Luisa Bernardinelli: conceptualisation (supporting), investigation; writing -- review
and editing. Both authors approved the final manuscript.

\bigskip

\section*{Funding}
This work received no specific funding.

\bigskip

\section*{Conflict of interest statement}
The authors declare no conflicts of interest.

\bigskip

\section*{Use of generative AI}
During the preparation of this manuscript the authors used generative AI to check
language and internal consistency. The authors take full responsibility for the
content of the publication.

\FloatBarrier
\clearpage
\appendix
\renewcommand{\thesection}{S\arabic{section}}
\setcounter{section}{0}
\renewcommand{\thetable}{S\arabic{table}}
\setcounter{table}{0}
\renewcommand{\thefigure}{S\arabic{figure}}
\setcounter{figure}{0}
\section*{Supporting Information}

\bigskip

\section{A target intervention identified from trial and observational evidence}
\label{supp:stability}

\subsection{Clinical question and available information}

Consider a hypothetical trial of an antihypertensive agent $A$ added to usual care. Let $T=1$ denote assignment to $A$ and $T=0$ usual care alone. The target variable $S$ describes a prespecified blood-pressure profile assessed before an outcome period, and $Y$ indicates a cardiovascular event during the subsequent five years. For simplicity, distinguish two specified profiles, controlled ($S=1$) and elevated ($S=0$), and assume complete target assessment and follow-up. The profiles must encode the features of pressure history relevant to the proposed intervention; a single reading would generally be insufficient.

The baseline variables $C$ include prior pressure history, renal function, vascular disease, smoking and background treatment. In the regimes without $A$ described below, $C$ is assumed sufficient to control common causes of $S$ and $Y$. Randomisation does not establish that assumption. Time-varying common causes or effects of earlier treatment could require a longitudinal formulation.\cite{dawidDidelez2010}

Suppose the available trial report gives target frequencies and outcome risks within strata of $C$, but records linking $S$ to $Y$ are unavailable. In one stratum, the controlled profile occurs in 20\% of the control arm and 80\% of the active arm, while cardiovascular risk is 9.2\% in both. An observational cohort receiving the same usual care, without $A$, supplies joint measurements of $C,S,Y$. In the corresponding stratum, risk is 6\% with the controlled profile and 10\% with the elevated profile. Table~\ref{tab:fusion-data} treats these hypothetical quantities as population values to separate identification from sampling uncertainty.

\begin{widetable}[tbp]
\small
\caption{Hypothetical information within one baseline stratum $C=c$.}
\label{tab:fusion-data}
\begin{tabularx}{\linewidth}{@{}L{32mm}Ycc@{}}
\toprule
Source & Arm or target profile & Controlled profile & Event risk \\
\midrule
Trial & Usual care, $T=0$ & 20\% & 9.2\% \\
Trial & Agent $A$, $T=1$ & 80\% & 9.2\% \\
Cohort, no $A$ & Controlled, $S=1$ & --- & 6\% \\
Cohort, no $A$ & Elevated, $S=0$ & --- & 10\% \\
\bottomrule
\end{tabularx}
\end{widetable}

\subsection{Interventions and stability}

Define two hypothetical interventions, $g_0$ and $g_1$, in the trial population. Neither
administers $A$. Within each baseline stratum, $g_a$, for $a=0,1$,
assigns the target profile independently of other
patient characteristics, using the profile distribution
observed in the same stratum of trial arm $a$. In the illustrated stratum,
the probability of the controlled profile therefore changes from 20\% to 80\%.
The interventions used to generate these profiles affect $Y$ only through $S$. 
This restriction concerns the target interventions; other effects of the drug
remain permitted.

Let $\sigma$ distinguish the cohort ($O$), the trial control arm ($R_0$) and
the target interventions. All four regimes exclude $A$. The required outcome stability is
\begin{equation}
\begin{gathered}
Y\indep\sigma\mid(S,C),\\
\sigma\in\{O,R_0,g_0,g_1\}.
\end{gathered}
\label{eq:fusion-stability}
\end{equation}
For the same baseline characteristics and target profile, the outcome distribution is assumed common across these regimes. The missing $\sigma\to Y$ arrow in Figure~\ref{fig:fusion}\subref{fig:fusion-no-agent} represents this claim. The indicator permits differences in the distributions of $C$ and $S$; the target interventions themselves use the trial's baseline distribution.

\begin{widefigure}[tbp]
\begin{subfigure}[t]{.48\linewidth}
\centering
\begin{tikzpicture}[x=.95cm,y=.95cm]
\node[cnode] (T) at (0,0) {$T$};
\node[cnode] (S) at (2.7,0) {$S$};
\node[cnode] (Y) at (5.4,0) {$Y$};
\node[cnode] (C) at (2.7,1.8) {$C$};
\draw[ar] (T)--(S);
\draw[ar] (S)--(Y);
\draw[ar] (C)--(S);
\draw[ar] (C)--(Y);
\draw[ar] (T) to[out=-36,in=-144] (Y);
\path (0,-1.2)--(5.4,-1.2);
\end{tikzpicture}
\caption{Randomised trial, regime $R$.}
\label{fig:fusion-trial}
\end{subfigure}\hfill
\begin{subfigure}[t]{.48\linewidth}
\centering
\begin{tikzpicture}[x=.95cm,y=.95cm]
\node[lnode] (F) at (0,1.8) {$\sigma$};
\node[cnode] (C) at (2.7,1.8) {$C$};
\node[cnode] (S) at (2.7,0) {$S$};
\node[cnode] (Y) at (5.4,0) {$Y$};
\draw[bd] (F)--(C);
\draw[bd] (F)--(S);
\draw[ar] (C)--(S);
\draw[ar] (C)--(Y);
\draw[ar] (S)--(Y);
\path (0,-1.2)--(5.4,-1.2);
\end{tikzpicture}
\caption{Regimes without agent $A$.}
\label{fig:fusion-no-agent}
\end{subfigure}
\caption{Graphs for the combination of evidence. (a) $T$ denotes randomised assignment, $S$ the target profile, $C$ baseline covariates and $Y$ the outcome. The direct $T\to Y$ edge permits effects bypassing $S$. (b) The non-stochastic regime indicator distinguishes the cohort, trial control arm and two target interventions. Its missing arrow into $Y$ asserts stability given $S,C$. That assumption does not require the same outcome relation when $A$ is administered. Dashed arrows permit regime dependence; causal arrows are solid.}
\label{fig:fusion}
\end{widefigure}

Stability requires adequate confounding control in the cohort, comparable meanings of $S,C,Y$ and no remaining differences in history or care that alter outcome risk at fixed $S,C$. The cohort must include the baseline strata and target profiles required by the interventions (positivity).\cite{dawidDidelez2010}

\subsection{Identification and numerical illustration}

Write
\begin{align*}
m_O(s,c)&=P_O(Y=1\mid S=s,C=c),\\
q_a(s\mid c)&=P_R(S=s\mid T=a,C=c),
\end{align*}
and let $\pi_R(c)$ denote the trial's baseline distribution. Under stability and positivity, the risk under $g_a$, for $a=0,1$, is
\begin{equation}
R(g_a)=\sum_c\pi_R(c)\sum_s m_O(s,c)q_a(s\mid c).
\label{eq:fusion-formula}
\end{equation}
The cohort supplies outcome risks at each target profile; the trial supplies profile frequencies. Averaging over $C$ puts the comparison in the trial population. Stability between $O$ and $R_0$ also implies that $R(g_0)$ equals the trial control risk.

In the illustrated stratum, the risks $R_c(g_a)$ are
\begin{align*}
R_c(g_0)&=0.20(0.06)+0.80(0.10)\\
        &=0.092,\\
R_c(g_1)&=0.80(0.06)+0.20(0.10)\\
        &=0.068.
\end{align*}
Thus the specified target intervention reduces risk by 2.4 percentage points:
\[
\psi_c=R_c(g_1)-R_c(g_0)=-0.024.
\]
Risk under $A$ is nevertheless 0.092, exceeding $R_c(g_1)$ by 0.024. This remaining contrast includes whatever additional actions or interactions distinguish administering $A$ from generating the same target distribution without it. It does not identify those processes separately.

The estimand is the effect of a stochastic target intervention, not a natural indirect effect assigning each patient the target value they would individually have under another treatment. Under the assumptions, a beneficial change in the target distribution has been identified in this patient group despite the null total trial effect.

With sampled data, uncertainty from both sources must be propagated and sensitivity to residual confounding and failures of stability assessed. Reproducing the observed trial control risk is a useful check but cannot establish stability for unobserved target interventions. If joint trial data were available, its control arm could supply the outcome relation under the same relevant assumptions. The cohort is useful here because the available trial summaries omit it.

\bigskip

\section{A second agent and bounds on additional effects}
\label{supp:bounds}

Let $p(r,j)$ be the risk of an adverse clinical outcome within a fixed follow-up time in the same population and care setting. Imagine choosing separately which agent to administer and which target profile to produce. The index $j$ specifies the agent: $j=0$ means neither agent, while $j=A$ or $j=B$ means the indicated agent. The index $r$ specifies the target profile: $r=1$ sets a prespecified response profile and $r=0$ its reference profile. In the CAST setting, these might represent a specified degree and duration of ectopy suppression and the corresponding profile without suppression. The intervention setting the target is assumed to affect the outcome only through that profile.

Define
\begin{align*}
\psi&=p(1,0)-p(0,0),\\
\delta_j&=p(1,j)-p(1,0).
\end{align*}
If a trial realises the comparison between $(1,j)$ and $(0,0)$, its risk difference is
\[
\theta_j=\psi+\delta_j.
\]
The additional effect $\delta_j$ is defined at the fixed target profile and includes interactions present there. Relating the identity to an actual trial requires delivery of that profile and a meaningful comparison between the specified interventions.

An overall effect near zero for $A$ is compatible with $(\psi,\delta_A)=(0,0)$ and with $(-b,b)$ for $b>0$. A trial of $B$ becomes informative about $\psi$ if independent evidence sufficiently constrains $\delta_B$. Suppose a fixed bound $|\delta_B|\leq\varepsilon$ is justified. A valid confidence interval $[l_B,u_B]$ for $\theta_B$ then yields the conservative confidence region
\begin{equation}
[l_B-\varepsilon,\ u_B+\varepsilon]
\label{eq:bound}
\end{equation}
for $\psi$. The coverage statement holds when the causal interpretation and bound are correct. If the bound is estimated, its uncertainty must also be incorporated; treating its estimate as fixed is insufficient.

For example, an interval $[-0.05,-0.03]$ and $\varepsilon=0.01$ give $[-0.06,-0.02]$, excluding zero. An interval $[-0.01,0.01]$ gives $[-0.02,0.02]$, excluding a target benefit of at least three percentage points while leaving smaller contributions unresolved. With $\varepsilon=0.04$, the regions become $[-0.09,0.01]$ and $[-0.05,0.05]$, and those conclusions no longer follow. All values are hypothetical.

The record should state the evidence supporting the bound
and assess how wider bounds change the conclusion.
Without a sufficiently narrow bound, a trial of $B$
may establish its clinical value while leaving the
target contribution unresolved.

\bigskip

\section{Genetic instruments in the lipid-lowering graph}
\label{supp:genetic}

\subsection{Markers, proxies and the latent background}

Observed markers or scores $G_H,G_P$ need not themselves be functional. The reduced proxy forks $G_j\leftarrow G_j^{*}\to H$ or $P$ follow the linkage representation discussed by Lawlor et al.; $G_j^{*}$ can denote several functional variants. The arrow into an observed marker represents its proxy relationship, not biological production of one allele by another.\cite{lawlor2008mr} Figure~\ref{fig:genetic-rivals} shows the latent background $U$ and the rival explanations of a genetic association.

\bigskip

\begin{widefigure}[tbp]
\begin{tikzpicture}[x=1cm,y=.8cm]
\node[bhhidden] (v) at (-2.6,2.2) {$V$\\[-1pt]{\scriptsize ancestry}};
\node[bhnode] (g) at (-2.6,0) {$G$};
\node[bhhidden] (gs) at (0,0) {$G^{*}$};
\node[bhhidden] (gss) at (0,-2.2) {$G^{**}$};
\node[bhtarget] (t) at (2.8,0) {$T$\\target};
\node[bhnode] (m) at (5.3,0) {$M$\\[-1pt]{\scriptsize $R,L,W$}};
\node[bhnode] (yy) at (7.8,0) {$Y$};
\node[bhu] (u) at (4.05,2.2) {$U$};
\draw[bha] (gs)--(g); \draw[bha] (gs)--(t); \draw[bha] (t)--(m); \draw[bha] (m)--(yy);
\draw[bhua] (u)--(t); \draw[bhua] (u)--(m); \draw[bhua] (u)--(yy);
\draw[bhriv] (v)--(gs);
\draw[bhriv] (v.north east) to[out=20,in=100] node[bhlab,pos=.5] {population structure} (yy.north);
\draw[bhriv] (gss)--(g);
\draw[bhriv] (gs.south east) to[out=-25,in=-140] node[bhlab,pos=.45] {pleiotropy} (yy.south west);
\draw[bhriv] (gss.east) to[out=-5,in=-80] node[bhlab,pos=.5] {linkage} (yy.south east);
\path (-3.1,-3.6)--(8.5,3.9);
\end{tikzpicture}
\caption{Latent background and rival explanations of a genetic association, for a
single target $T$ whose intermediates $M$ stand for $R$, $L$ and $W$ of
Figure~\ref{fig:statins-rivals}. The latent vector $U$ makes $T$ a collider on the path
from $G^{*}$ to $Y$, so that conditioning on the target or its descendants opens a path
through $U$. Dashed red arrows show three rival explanations of a marker--outcome
association: population structure $V$, a second functional variant $G^{**}$ in linkage
with the marker $G$, and pleiotropy of $G^{*}$. The exclusions in
Figure~\ref{fig:statins-rivals} assume all three absent.
The main graph assumes that adjustment for $C$
adequately controls population structure and that
the pleiotropic and linkage pathways shown here
are absent.}
\label{fig:genetic-rivals}
\end{widefigure}

\bigskip

Under the instrument assumptions, a genetic association
with either outcome supports an effect of target
perturbation but does not identify the mediating pathway.
Exclusion of effects outside the target is weaker than
exclusion of effects outside LDL. Adjusting for measured
intermediates does not resolve this distinction:
conditioning on them can open paths through $U$
(Figure~\ref{fig:genetic-rivals}).

\subsection{Comparability of the two genetic perturbations}

The comparison of the two genetic perturbations in Section~\ref{sec:statins} rests on a stability restriction of its own. Let $\tau$ distinguish HMGCR from PCSK9 perturbation. In the submodel of panel~(a) without bypass edges from $H$, $P$ and $R$, and with no drug assigned, comparability requires $Y_C\indep\tau\mid(L,U,C)$: the coronary-outcome kernel given the LDL profile and $U,C$ is shared by the two perturbations. This restriction does not by itself imply equal coronary associations per unit of LDL, which is why Section~\ref{sec:statins} adds a quantitative working assumption of common proportionality. Together with that assumption, the restriction supports interpreting the concordance of the two scores. Complementary evidence can thus inform causal explanations and further comparisons without requiring the reported genetic and trial estimates to be pooled.

\subsection{Adiposity}

The adiposity findings concern the edges into $W$ in panel (b). Ference
and colleagues found associations with bodyweight and other adiposity
measures for the HMGCR score and none for the PCSK9 score. \cite{ference2016pcsk9}
Schmidt and colleagues, using four PCSK9 variants, reported higher
bodyweight (1.03 kg; 95\% CI 0.24--1.82 per mmol/L lower LDL) and
waist-to-hip ratio, with no statistically detectable association with
body-mass index.\cite{schmidt2017pcsk9} These findings leave
uncertainty about the size
of the effect of PCSK9 perturbation on adiposity. If $P\to W$ is weak,
the similar diabetes associations of the two targets weigh against weight
gain as the main shared explanation; if it is present, weight remains
a candidate. In either case, weight mediation requires a causal
effect of adiposity on diabetes, either directly
or through downstream processes such as changes in LDL.

\bigskip

\section{Tipifarnib and an alternative route to target function}
\label{supp:tipifarnib}

Tipifarnib was investigated as an addition to gemcitabine chemotherapy in
patients with advanced pancreatic cancer, with the aim of improving
survival.\cite{vancutsem2004tipifarnib} One biological rationale concerned
KRAS, a protein that acts as a molecular switch in signalling pathways
regulating cell growth and survival. Activating mutations in the \emph{KRAS}
gene are common in pancreatic ductal adenocarcinoma and can keep
these growth-promoting signals active.

KRAS requires association with cell membranes to perform its signalling
function. Farnesyltransferase attaches a lipid group that supports this 
association, and tipifarnib inhibits that enzyme. The KRAS-based rationale
therefore required enzyme inhibition to reduce functional,
membrane-associated KRAS sufficiently to affect tumour growth.
However, experiments published before the pancreatic phase~III trial
showed that KRAS could undergo an alternative lipid modification,
geranylgeranylation, during farnesyltransferase inhibition, allowing
membrane association to persist.\cite{james1996kras,whyte1997kras,
rowell1997kras,lerner1997kras}
Figure~\ref{fig:tipifarnib} represents this alternative route.

This reconstruction identifies a delivery question;
it does not establish the investigators' beliefs or
show that a causal record would have changed their decisions.

\begin{widefigure}[tbp]
\begin{tikzpicture}[x=1cm,y=1cm,
 n/.style={cnode,text width=19mm,font=\sffamily\small},
 nw/.style={cnode,text width=25mm,font=\sffamily\small},
 es/.style={font=\small,fill=white,inner sep=1pt}]
\node[n] (A) at (0,0) {Tipifarnib};
\node[n] (B) at (2.7,0) {FTase\\ blockade};
\node[n] (KF) at (5.7,1.7) {Farnesylated\\ KRAS};
\node[nw] (KG) at (5.7,-1.7) {Geranyl-\\ geranylated\\ KRAS};
\node[n] (G) at (2.7,-3.5) {GGTase-I\\ activity};
\node[nw] (M) at (9,0) {Membrane-\\ localised\\ KRAS};
\node[n] (R) at (12,0) {RAS\\ signalling};
\node[n] (Y) at (12,-2.7) {Progression\\ or death};
\draw[ar] (A)--node[es,above] {$+$}(B);
\draw[ar] (B)--node[es,above left] {$-$}(KF);
\draw[ar] (B)--node[es,below left] {$+$}(KG);
\draw[ar] (G)--node[es,below right] {$+$}(KG);
\draw[ar] (KF)--node[es,above right] {$+$}(M);
\draw[ar] (KG)--node[es,below right] {$+$}(M);
\draw[ar] (M)--node[es,above] {$+$}(R);
\draw[ar] (R)--node[es,right] {$+$}(Y);
\end{tikzpicture}
\caption{A delivery question raised by alternative prenylation.
FTase denotes farnesyltransferase. Its inhibition reduces farnesylated
KRAS but permits compensatory geranylgeranylation while GGTase-I
remains active. Both forms can support membrane localisation.
Signs express the proposed directions, not their magnitudes or
their adequacy at a particular dose in human tumours.}
\label{fig:tipifarnib}
\end{widefigure}

A tumour pharmacodynamic study could have tested whether
the intended exposure reduced functional,
membrane-associated KRAS. The absence of a survival
benefit with added tipifarnib did not by itself
distinguish inadequate target modification from an
ineffective or offset clinical contribution.
\cite{vancutsem2004tipifarnib}

\FloatBarrier

\end{document}